\documentclass{article}  
\usepackage{lajolla2006}
\usepackage{graphicx}
\usepackage{epsfig,latexsym}
\usepackage{amssymb}

\newcommand{\be}{\begin{eqnarray}}
\newcommand{\ee}{\end{eqnarray}}

\frompage{000} \topage{000}                                              
\def\rightmark{Progress in lattice QCD at finite temperature}
\def\leftmark{P. Petreczky }

\usepackage{graphicx}
\def\be{\begin{equation}}
\def\ee{\end{equation}}

\newcommand{\beq}{\begin{equation}}
\newcommand{\eeq}{\end{equation}}
\newcommand{\ber}{\begin{eqnarray}}
\newcommand{\eer}{\end{eqnarray}}
\newcommand{\berr}{\begin{eqnarray*}}
\newcommand{\eerr}{\end{eqnarray*}}

\title{Progress in lattice QCD at finite temperature }
\authors{
{P\'eter Petreczky
}\\[2.812mm]
{\normalsize
Physics Deparment and RIKEN-BNL Research Center, Brookhaven National Laboratory, Upton, NY 11973 
}}
 
\abstract{I review current status of lattice QCD calculations of the deconfining
transition at finite temperature and quarkonia spectral functions.
}


\begin{document}
 
\maketitle
\setcounter{page}{1}


\section{Introduction}
\label{intro}

One of the most interesting properties of QCD is the existence
of the deconfining transition at some temperature $T_c$, above 
which hadrons no longer exist and the strongly interacting matter
is described in terms of quarks and gluons. Therefore an important
task for lattice QCD is to study the nature of this transition and 
to determine the precise value of the transition temperature.
In the next section I will discuss this problem more in detail.

Although light hadrons cannot exist above the deconfinement temperature 
the situation can be different for heavy quarkonia.
Because of the heavy quark mass quarkonia binding can be understood
in terms of the static potential. 
General considerations suggest  that quarkonia could melt at
temperatures above the deconfinement temperature
as a result of modification of inter-quark forces (color screening).
It has been conjectured by that melting of
different quarkonia states due to color screening 
can signal Quark Gluon Plasma formation in heavy ion collisions \cite{MS86}.
Many studies  of quarkonia dissolution 
rely heavily on potential models 
\cite{karsch88,digal01a,digal01b,shuryak04,wong04,alberico}.
However it is very unclear if such models are valid 
at finite temperature  \cite{petreczkyhard04}. 

The problem of quarkonium dissolution can be studied more 
rigorously in terms of meson (quarkonium) spectral functions.
Lattice calculation of charmonium spectral functions
appeared recently and suggested, contrary to potential models,
that $J/\psi$ and $\eta_c$ survive at temperatures as high as $1.6T_c$ 
\cite{umeda02,asakawa04,dattalat02,datta04}. It has been also found that $\chi_c$ 
melts at temperature of about  $1.1T_c$ \cite{dattalat02,datta04,jhw05}. 
There are also preliminary calculations of the bottomonium
spectral functions \cite{lat05,panic05}. In section 3
I will discuss charmonium correlators and spectral functions. Finally
conclusions will be given in the last section. 

\section{The finite temperature transition in QCD}

We would like to know  what is
the nature of the transition to the new state of mater and
what is the temperature where it happens
\footnote{I will talk here about the QCD finite temperature transition
irrespective whether it is a true phase transition or a crossover
and $T_c$ will always refer to the corresponding temperature.}.
In the case of QCD without dynamical quarks, i.e. SU(3) gauge theory
these questions have been answered. It is well
established that the phase transition is 1st order \cite{fukugita89}.
Using standard and improved actions the corresponding transition
temperature was estimated to be $T_c/\sqrt{\sigma}=0.632(2)$
\cite{edwin} ($\sigma$ is the string tension).
The situation for QCD with dynamical quarks is much more difficult.
Not only because the inclusion of dynamical quarks increases the
computational costs by at least two orders of magnitude but also
because the transition is very sensitive to the quark masses.
Conventional lattice fermion formulations break global
symmetries of continuum QCD (e.g. staggered fermion
violate the flavor symmetry) which also introduces additional complications.
Current lattice calculations suggest
that transition in QCD for physical quark masses is not a
true phase transition but a crossover
\cite{karsch01,frithjoflat03,fodor04,hyp,milcthermo}.
The transition appears to be first order only for very small
quark masses corresponding to pion mass of about $67$MeV
for three degenerate flavors \cite{frithjoflat03}. At finite temperature 
there is only one transition in QCD. The deconfinement transition
identified with rapid increase in the degrees of freedom or 
increase of the Polyakov loop \cite{petrov04} coincides with
the chiral transition \cite{karsch01}.

\begin{figure}
\centerline{
\includegraphics[width=9cm]{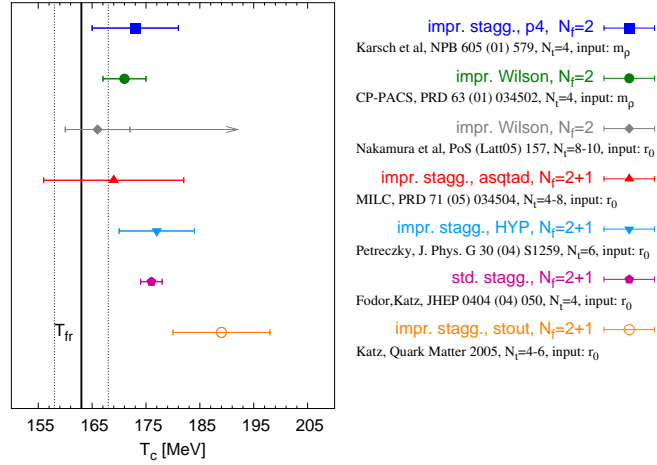}
\vspace*{-0.2cm}
}
\caption{Summary of lattice results on the transition
temperature $T_c$ taken from Refs.
\cite{frithjoflat03,fodor04,hyp,milcthermo,cppacsnf2,nakamuralat05}
as well as unpublished result from S. Katz presented on Quark Matter 2005.
The horizontal line and the band is the chemical freezout temperature
at the highest RHIC energy \cite{starwhite}.
\label{fig:tc}}
\end{figure}

Recent lattice results for the transition temperature $T_c$ from
Wilson fermions \cite{cppacsnf2,nakamuralat05},
improved \cite{karsch01,hyp,milcthermo} and unimproved staggered fermions
\cite{fodor04} with 2 and 2+1 flavors of dynamical quarks
are summarized in Fig. \ref{fig:tc}. The errors shown
in Fig. \ref{fig:tc} are only statistical with the exception
of the data point from the MILC collaboration,
where the large error partly comes from the continuum
extrapolation and also includes systematic error in scale setting.
One should note that calculations with Wilson fermions are done at
very large quark masses (of the order of the strange quark mass
or larger) and therefore the uncertainty in the value of $T_c$ is dominated
by the uncertainty of the extrapolation in the quark mass \cite{nakamuralat05} and shown
in  Fig. \ref{fig:tc} as an horizontal arrow. In the figure I also show
the chemical freeze-out temperature at the highest RHIC energy \cite{starwhite}.
Since the ``critical'' energy density $\epsilon_c=\epsilon(T_c)$
(i.e. the energy density at
the transition) scales as $T_c^4$ the error in $T_c$ is the dominant source
of error in $\epsilon_c$ \cite{petreczkylat04}.

\section{Charmonia correlators and spectral functions}%

In  lattice QCD we calculate correlators of point
meson operators of the form 
\begin{equation}
	J_H(t,x)=\bar q(t,x) \Gamma_H q(t,x),
\end{equation}
where $\Gamma_H=1,\gamma_5, \gamma_{\mu}, \gamma_5 \gamma_{\mu}, 
\gamma_{\mu} \gamma_{\nu}$ 
and fixes the quantum number of the channel to 
scalar, pseudo-scalar, vector, axial-vector and tensor channels correspondingly.
The relation of these quantum number channels to different meson states is given
in Tab. \ref{tab.channels}.
Most dynamic properties of a finite temperature system are incorporated 
in the spectral function. The spectral function $\sigma(p_0,\vec{p},T)$ for a given 
mesonic channel $H$ in a system at temperature $T$ can be defined 
through the Fourier transform of the real time two-point functions
$D^{>}(x_0,\vec{x},T)$ and $D^{<}(x_0,\vec{x},T)$ or, equivalently, as the imaginary part of 
the Fourier transformed retarded 
correlation function \cite{lebellac}.
The Euclidean time correlator calculated on the lattice
\beq
G_H(\tau, \vec{p},T) = \int d^3x e^{i \vec{p} \cdot \vec{x}} 
\langle T_{\tau} J_H(\tau, \vec{x}) J_H(0,
\vec{0}) \rangle
\eeq
is an analytic continuation of the real time correlator
$G_H(\tau,\vec{p},T)=D^{>}(-i\tau,\vec{p},T)$. 
Using this fact and the  Kubo-Martin-Schwinger
(KMS) condition \cite{lebellac} for the correlators
$D^{>}_H(x_0,\vec{x},T)=D^{<}(x_0+i/T,\vec{x},T)$,
one can relate the Euclidean propagator $G_H(\tau,\vec{p},T)$ to the 
spectral function through the integral
representation
\ber
G(\tau, \vec{p},T) &=& \int_0^{\infty} d \omega
\sigma(\omega,\vec{p},T) K(\omega, \tau,T) \label{eq.spect} \nonumber\\
K(\omega, \tau,T) &=& \frac{\cosh(\omega(\tau-1/2
T))}{\sinh(\omega/2 T)}.
\label{eq.kernel}
\eer
\begin{table}
\begin{tabular}
[c]{||c|c|c||c|}\hline
$\Gamma$ & $^{2S+1}L_{J}$ & $J^{PC}$ & $u\overline{u}$\\\hline
$\gamma_{5}$ & $^{1}S_{0}$ & $0^{-+}$ & $\pi$\\
$\gamma_{s}$ & $^{3}S_{1}$ & $1^{--}$ & $\rho$\\
$\gamma_{s}\gamma_{s^{\prime}}$ & $^{1}P_{1}$ & $1^{+-}$ & $b_{1}$\\
$1$ & $^{3}P_{0}$ & $0^{++}$ & $a_{0}$\\
$\gamma_{5}\gamma_{s}$ & $^{3}P_{1}$ & $1^{++}$ & $a_{1}$\\
\hline
\end{tabular}%
\begin{tabular}
[c]{|cc|}\hline
$c\overline{c}(n=1)$ & $c\overline{c}(n=2)$\\\hline
$\eta_{c}$ & $\eta_{c}^{^{\prime}}$\\
$J/\psi$ & $\psi^{\prime}$\\
$h_{c}$ & \\
$\chi_{c0}$ & \\
$\chi_{c1}$ & \\
\hline
\end{tabular}
\begin{tabular}[c]{|cc|}\hline
$b\overline{b}(n=1)$ & $b\overline{b}(n=2)$\\
\hline
$\eta_b$ & $\eta_b'$ \\
$\Upsilon(1S)$ & $\Upsilon(2S)$\\
$h_b$ & \\
$\chi_{b0}(1P)$& $\chi_{b0}(2S)$\\
$\chi_{b1}(1P)$& $\chi_{b1}(2P)$\\
\hline
\end{tabular}
\caption{Meson states in different channels
\label{tab.channels}}
\end{table}
To reconstruct the spectral function from the lattice correlator 
$G(\tau,T)$ this integral representation should be inverted. 
Since the number of data points is less than the number of degrees
of freedom (which is ${\cal O}(100)$ for reasonable discretization of
the integral ) spectral functions can be reconstructed only using the
Maximum Entropy Method (MEM) \cite{asakawa01}. 
In this method one looks for a spectral function which
maximizes the
conditional probability $P[\sigma|DH]$ of having the spectral function $\sigma$ given
the data $D$ and some prior knowledge $H$ which for positive definite spectral function
can be written as 
\beq
 P[\sigma|DH]=\exp(-\frac{1}{2} \chi^2 + \alpha S),
\label{eq:PDH}
\eeq
where 
$
S=\int d \omega [ \sigma(\omega)-m(\omega)-\sigma(\omega)
  \ln(\sigma(\omega)/m(\omega)) ]
$
is the Shannon - Janes entropy. 
The real function $m(\omega)$ is called the default model and parametrizes all
additional prior knowledge about the
spectral functions, such as the asymptotic behavior at high energy  \cite{asakawa01}.
In order to have sufficient
number of data points either very fine isotropic lattices \cite{dattalat02,datta04,panic05}
or anisotropic lattices \cite{umeda02,asakawa04,jhw05,lat05} have been used.

\begin{figure}
	\includegraphics[width=6cm]{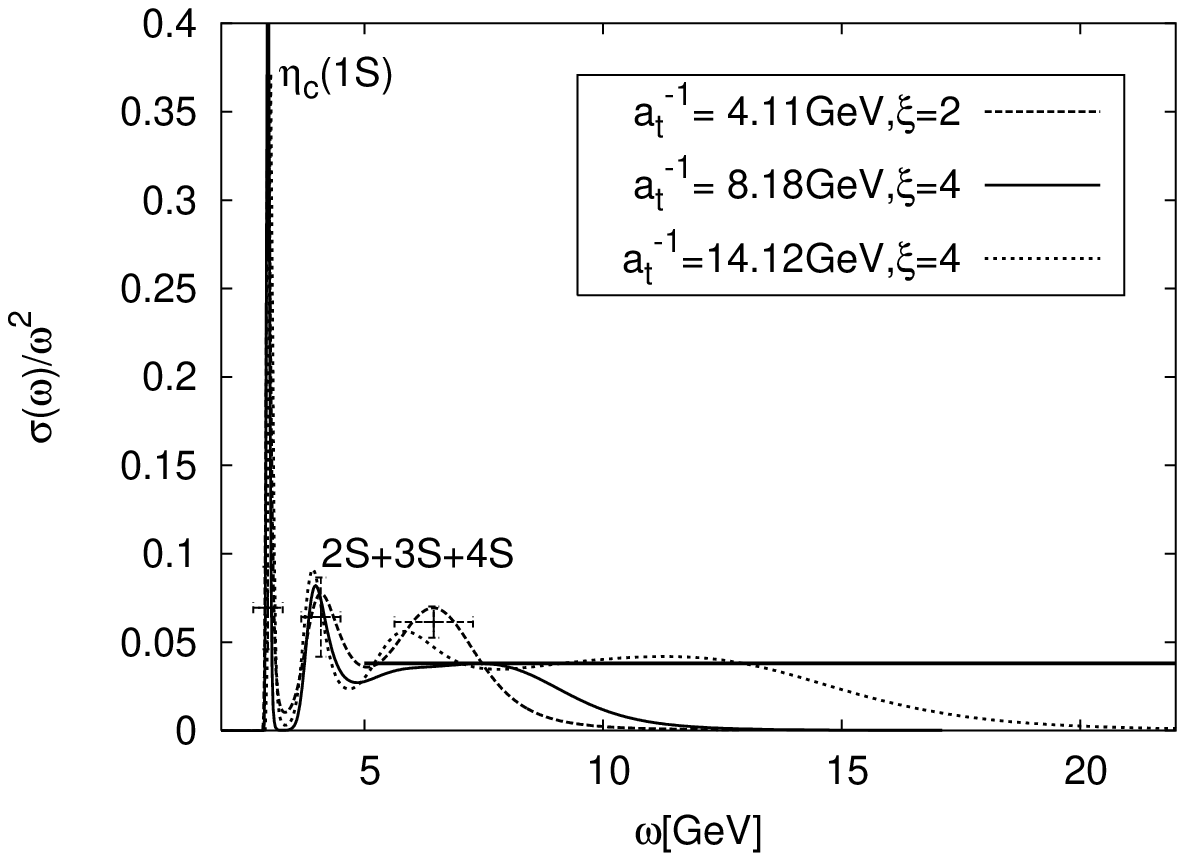}
	\includegraphics[width=6cm]{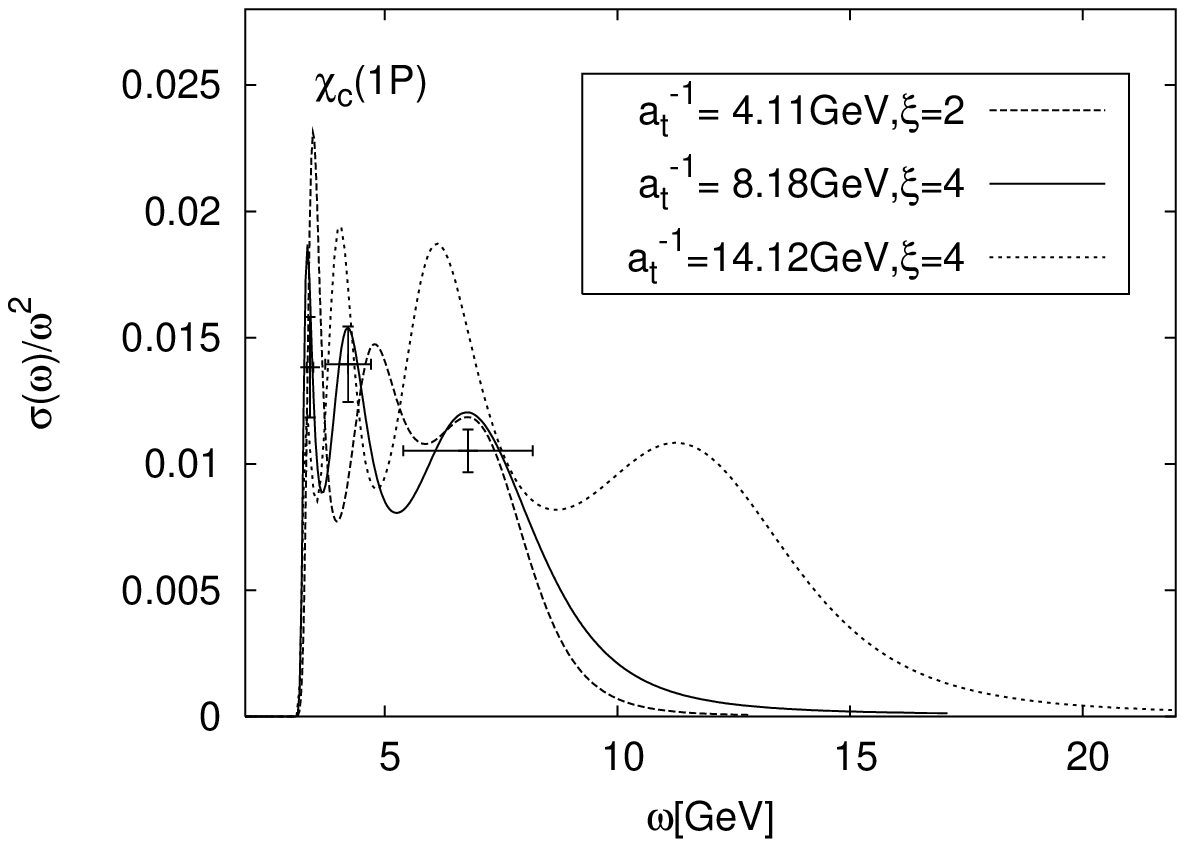}
\vspace*{-0.3cm}
\caption{
Charmonium spectral function in  the pseudo-scalar channel (left) and 
the scalar channel (right) at different lattice spacings and zero temperature from Ref. \cite{jhw05}. }
\label{spfT0}
\vspace*{-0.3cm} 
\end{figure}

The spectral function for pseudo-scalar charmonium spectral functions calculated on 
anisotropic lattice \cite{jhw05} is shown in Fig. \ref{spfT0}.
The first peak in the spectral function corresponds to $\eta_c(1S)$ state.  The position of the
peak and the corresponding amplitude (i.e. the area under the peak) are in good agreement
with the results of simple exponential fit. The second peak in the spectral function is most likely
the combination of several excited states as its position and amplitude is higher than what one would
expect for pure 2S state. The spectral function becomes sensitive to the effects of the finite lattice spacings
for $\omega>5$GeV. In this $\omega$ region the spectral functions becomes also sensitive to the choice
of the default model.  Also shown in Fig. \ref{spfT0}  is the spectral function in the
scalar channel from Ref. \cite{jhw05}. 
The 1st peak corresponds to $\chi_{c0}(1P)$ state. The correlator is more noisy
in the scalar channel than in the pseudo-scalar one. As the results the $\chi_{c0}(1P)$ peak
is less pronounced and has larger statistical errors. The peak position and the area under the peak
is consistent with the simple exponential fit. 
As in the pseudo-scalar case individual excited states are
not resolved and the spectral function depends on the lattice spacing and default model for $\omega>5$GeV. 
Similar results have been found for the vector and axial-vector
channels which correspond to $J/\psi$ and $\chi_{c1}$ states respectively.

\begin{figure}
\includegraphics[width=6cm]{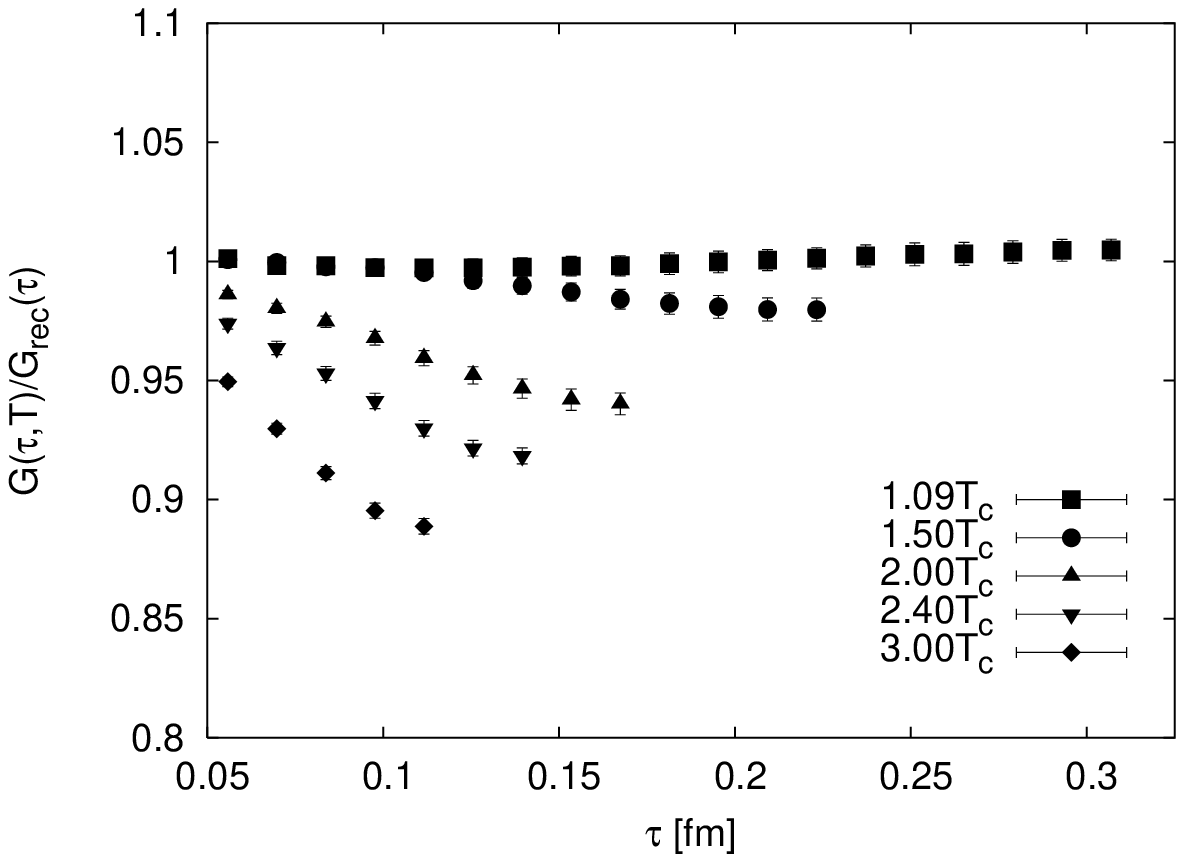}
\includegraphics[width=6cm]{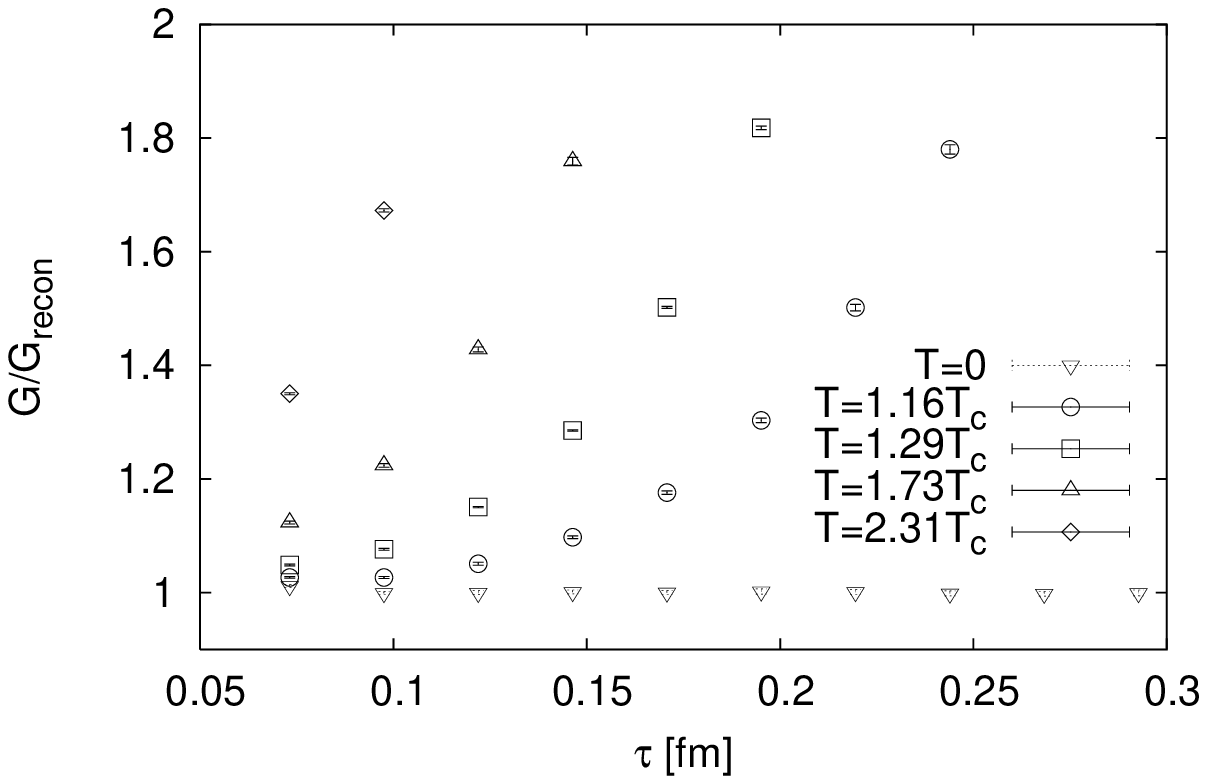}
\vspace*{-0.3cm}   
\caption{The ratio $G(\tau,T)/G_{recon}(\tau,T)$ of charmonium for pseudoscalar channel at  at $a_t^{-2}=14.11$GeV
(left) and scalar channel at  at $a_t^{-2}=8.18$GeV  (right) at different
temperatures \cite{jhw05}.}
\label{ratioc}
\vspace*{-0.3cm} 
\end{figure}

We would like to know what happens to different charmonia states
at temperatures above the deconfinement temperature $T_c$. With
increasing temperature it becomes more and more difficult to
reconstruct the spectral functions as both the number of available
data points as well as the physical extent of the time direction
(which is $1/T$) decreases. Therefore it is useful to study the
temperature dependence of charmonia correlators first. From
Eq. (\ref{eq.kernel}) it is clear that the temperature dependence 
of charmonia correlators come from two sources: the temperature
dependence of the spectral function and the temperature dependence of
the integration kernel $K(\tau,\omega,T)$. To separate out the 
trivial temperature dependence due to the integration kernel,
following Ref. \cite{datta04} for each temperature we calculate
the so-called reconstructed correlator  
$
G_{recon}(\tau,T)=\int_{0}^{\infty}d\omega
\sigma(\omega,T=0)K(\tau,\omega,T).
$
Now if we assume that there is no temperature dependence 
in the spectral function, then the ratio of the original and 
the reconstructed correlator should be close to one,
$G(\tau,T)/G_{recon}(\tau,T) \sim 1$. 
This way we can identify large changes in the  spectral function.
This gives reliable information about the fate of charmonia states above
deconfinement. 
In Fig.~\ref{ratioc} we show this ratio for pseudo-scalar and scalar 
channels correspondingly calculated on anisotropic lattice \cite{jhw05}. 
From the figures one can see that the pseudo-scalar correlators shows
only very small changes till $1.5T_c$ indicating that the $\eta_c$ state survives
till this temperature with little modification of its properties. On the other hand the scalar
correlator shows large changes already at $1.16T_c$ suggesting strong modification or
dissolution of the  $\chi_{c0}$ state at this temperature.

\begin{figure}
	\includegraphics[width=6cm]{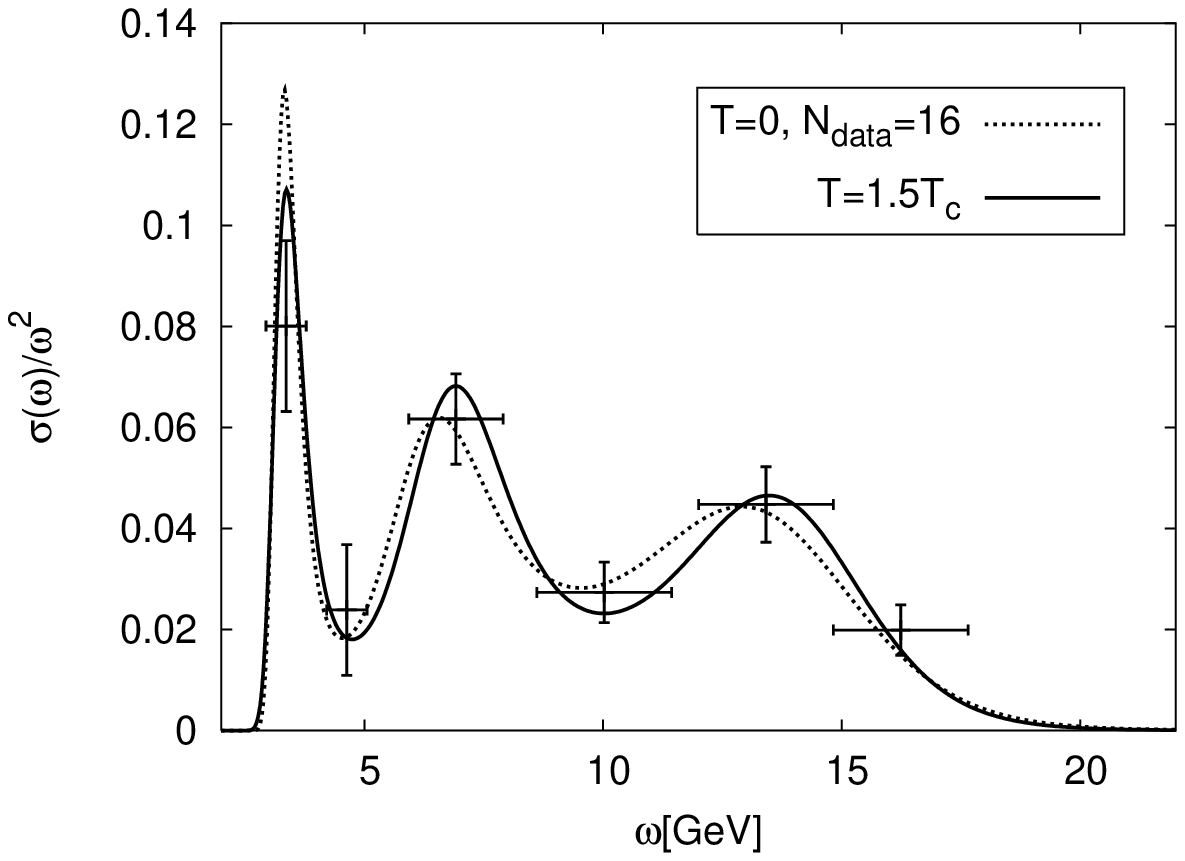}
	\includegraphics[width=6cm]{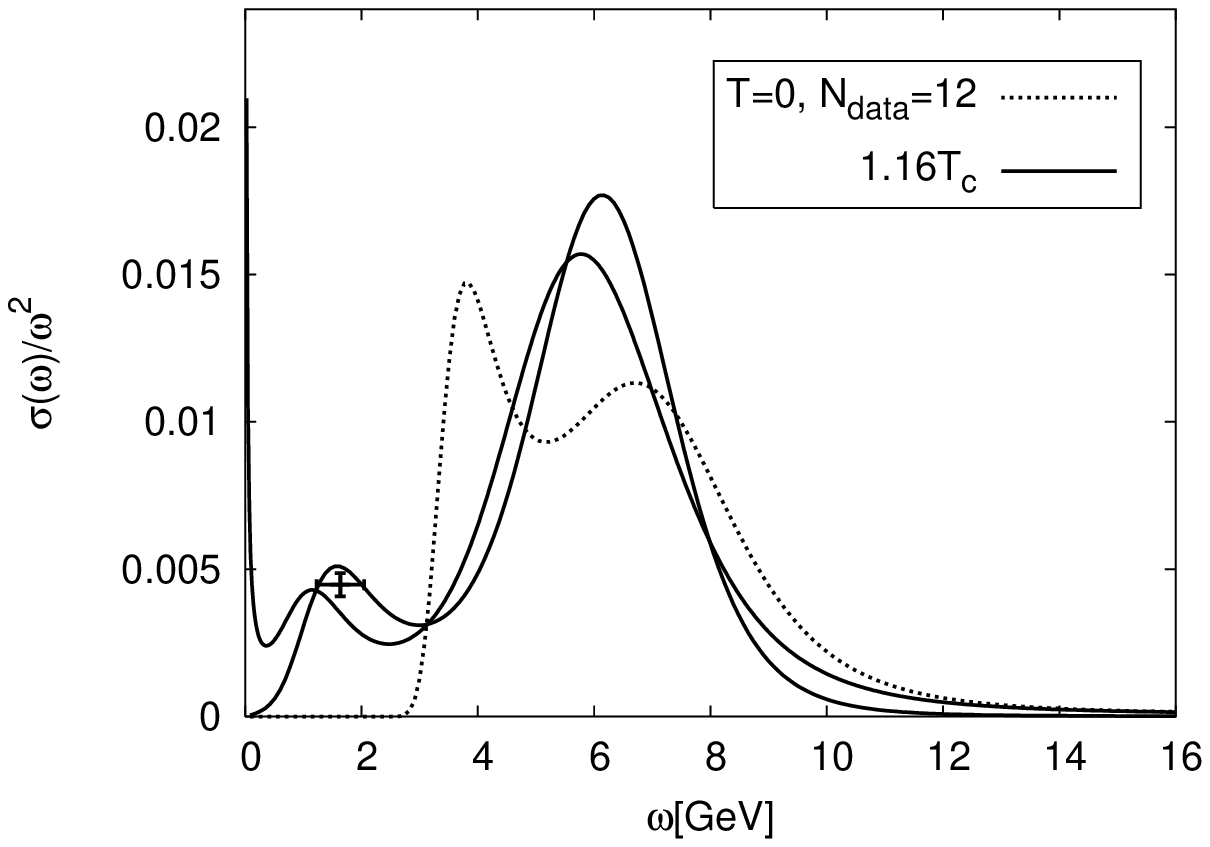}
\vspace*{-0.3cm}
\caption{
Charmonia spectral function in  the psuedoscalar channel at $a_t^{-2}=14.11$GeV  (left) and 
the scalar channel (right)   at $a_t^{-2}=8.18$GeV  for zero and finite temperature \cite{jhw05}. 
For finite temperature 
scalar channel two different default  models are shown.}
\label{fig.spfct}
\vspace*{-0.3cm} 
\end{figure}

More detailed information on different charmonia
states at finite temperature can be obtained by
calculating spectral functions using MEM.
The results of these calculations are show in
Figs.~\ref{fig.spfct}. 
Because at high temperature the temporal extent and the number of data
points where the correlators are calculated become smaller the spectral functions reconstructed
using MEM are less reliable.  To take into account possible systematic 
effects when studying  the temperature modifications of the spectral functions
we compare the finite temperature spectral functions against the zero temperature spectral functions
obtained from the correlator  using the same time interval and number of data points as available
at finite temperature. 
We see that spectral function in the pseudo-scalar channel show no temperature dependence
within the statistical errors.
This is in accord with the analysis of the correlation functions.
Also the spectral functions show very little dependence on the default model. 
Similar conclusion has been made in Ref. \cite{dattalat02,datta04} where correlators and
spectral functions have been calculated on very fine isotropic lattices as well as in
Ref. \cite{asakawa04} where anisotropic lattice have been used. The pseudo-scalar spectral function
was found to be temperature independent also in Ref. \cite{umeda02} where 
correlators of 
extended meson operators have been studied on anisotropic lattices. The study of the charmonium correlators
with different spatial boundary conditions provides further evidence for
survival of the $1S$ charmonia states well above the deconfinement transition
temperature \cite{iida}.

The scalar spectral function shows large changes at
$1.16T_c$ which is consistent with correlator-based analysis. Also the default model dependence
of the scalar correlator is large above the deconfinement transition (c.f. Fig. 3, right).
This means that the $\chi_{c0}$ ($^3P_0$) dissolves at this temperature. 
Similar results for the scalar spectral function have been reported in \cite{dattalat02,datta04}.
The results for the axial correlators and spectral functions are similar to scalar ones 
\cite{datta04} as expected. 

The vector correlator, however, has temperature dependence different
from that of the pseudo-scalar channel \cite{mylat05}. 
This is due to the fact that the vector current is
conserved and there is a contribution to the spectral functions at very small 
energy $\omega \simeq 0$
corresponding to heavy quark transport \cite{derek,mocsyhard04}.
The transport peak in the spectral functions can be written as \cite{derek}
\begin{equation}
\sigma_{low}(\omega)=\chi_s(T) \frac{T}{M} \frac{1}{\pi}\frac{\omega \eta}{\omega^2+\eta^2},
\end{equation}
where $\eta=T/M/D$ with $D$ being the heavy
quark diffusion constant. Furthermore $\chi_s(T)$ is the charm or beauty 
susceptibility and $M$ is the heavy quark mass.
To the first approximation the transport contribution to the spectral
function gives rise to a positive constant contribution to the correlator 
$G_{low}(\tau) \simeq \chi_s(T) T^2/M$ \cite{derek}, resulting in the  enhancement 
of the finite temperature correlators relative to the zero temperature ones, in agreement with the lattice 
data presented in \cite{mylat05,dattasewm04}. Finite value of the diffusion
constant $D$ will give rise to some curvature in $G_{low}(\tau)$. The smaller
the value of $D$ is,  the larger is the curvature in $G_{low}(\tau)$. Thus extracting
$G_{low}(\tau)$ from lattice data and estimating its curvature can give an estimate
for $D$.
This, however, requires very precise lattice data which are not yet available \cite{derek}.

\section{Conclusions}
In this contribution I discussed the status of finite temperature lattice QCD
calculations.  The transition in QCD appears to be a crossover and not a true
phase transition. The value of the transition temperature, however, is not well
know. Current lattice calculations give estimates between $167$ and $188$ MeV
with large statistical and systematic errors.  Recently charmonium correlators
and spectral functions have been calculated in lattice QCD and show that
$1S$ charmonia states survive above the transition temperatures till $1.6T_c$ while
the $1P$ state dissolve at $T \simeq 1.1T_c$.  Furthermore, the properties of
the $1S$ charmonia states are not modified siginificantly above $T_c$. On the
other hand potential model predict strong modification of charmonia properties
and therefore are not consistent with the lattice data on the correlators presented
above \cite{mocsyhard04}.

\section{Acknowledgments} 
This work was supported by U.S. Department of Energy under 
Contract No. DE-AC02-98CH10886.

\end{document}